\documentstyle[a4,12pt,epsf]{article}

\begin{document}

\baselineskip=18.8pt plus 0.2pt minus 0.1pt

\font\mybb=msbm10 at 12pt
\def\bb#1{\hbox{\mybb#1}}
\def\Z {\bb{Z}}
\def\R {\bb{R}}
\def\C {\bb{C}}
\def\J {\bb{J}}
\def\I {\bb{I}}
\def\CP {\bb{P}}

\font\mycc=msbm10 at 8pt
\def\cc#1{\hbox{\mycc#1}}
\def\sZ {\cc{Z}}
\def\sR {\cc{R}}
\def\sC {\cc{C}}

\def\ele{\mathop{\rm ele}\nolimits}
\def\mag{\mathop{\rm mag}\nolimits}
\def\tr{\mathop{\rm tr}\nolimits}
\def\Im{\mathop{\rm Im}\nolimits}
\def\diag{\mathop{\rm diag}\nolimits}
\def\rank{\mathop{\rm rank}\nolimits}
\def\Tr{\mathop{\rm Tr}\nolimits}
\def\mod{\mathop{\rm mod}\nolimits}
\def\ord{\mathop{\rm ord}\nolimits}

\renewcommand{\theparagraph}{(\roman{paragraph})}
\renewcommand{\thefootnote}{\fnsymbol{footnote}}
\renewcommand{\theequation}{\thesection.\arabic{equation}}
\makeatletter

\@addtoreset{equation}{section}
\renewcommand{\thefootnote}{\fnsymbol{footnote}}

\newcommand{\rf}[1]{(\ref{#1})}
\newcommand{\nn}{\nonumber}
\newcommand{\CR}{\nonumber \\}

\def\be{\begin{equation}}
\def\ee{\end{equation}}
\def\bea{\begin{eqnarray}}
\def\eea{\end{eqnarray}}

\def\pt{\partial}

\makeatother

\newcommand{\vs}{\vspace*}
\newcommand{\hs}{\hspace*}
\newcommand{\wt}{\widetilde}
\newcommand{\ol}{\overline}
\newcommand{\ul}{\underline}
\newcommand{\ra}{\rightarrow}
\newcommand{\lra}{\leftrightarrow}
\newcommand{\mms}[1]{\makebox[4ex]{$#1$}}
\newcommand{\sq}{\sqrt{2}\,}
\newcommand{\VEV}[1]{\left\langle #1\right\rangle}
\newcommand{\norm}[1]{\parallel #1\parallel}
\newcommand{\N}{{\cal N}}

\begin{titlepage}
\title{
\hfill\parbox{4cm}
{\normalsize KUNS-1543\\HE(TH)~98/17\\
{\tt hep-th/9811087}}\\
\vspace{1cm}
M-theory description of 1/4 BPS states \\ in ${\cal N}=4$ 
supersymmetric Yang-Mills theory 
}
\author{
Naoki {\sc Sasakura}\thanks{\tt sasakura@gauge.scphys.kyoto-u.ac.jp}
and Shigeki {\sc Sugimoto}\thanks{\tt sugimoto@gauge.scphys.kyoto-u.ac.jp}
\\[7pt]
{\it Department of Physics, Kyoto University, Kyoto 606-8502, Japan}
}

\date{\normalsize November, 1998}
\maketitle
\thispagestyle{empty}

\begin{abstract} 
We discuss BPS states preserving 1/4 supersymmetries of ${\cal N}=4$
supersymmetric Yang-Mills theory as M2-branes holomorphically
embedded and ending on M5-branes.
We use techniques in electrodynamics
to find the M2-brane configurations, and give
some explicit examples.
In case the M2-brane worldsheet has handles, the worldsheet moduli of
the M2-brane is constrained in a discrete manner.
Several aspects of multi-pronged strings in type IIB string theory
are beautifully reproduced in the M-theory description.
We also discuss the relation between the above construction
and the D2-brane dynamics in type IIA string theory.

\end{abstract}
\end{titlepage}

\section{Introduction}

The recent developments of non-perturbative string theory have provided
new powerful tools for analysis of non-perturbative properties of 
supersymmetric gauge theories.
A supersymmetric gauge theory can be studied as a low energy
effective field theory on a brane, and its BPS state may correspond to
a BPS configuration of a brane ending on the background brane in string theory.

The four-dimensional ${\cal N}=4$ $SU(N)$ SYM 
theory in Coulomb phase can be studied as 
the effective field theory on nearly coincident $N$ parallel D3-branes
in type IIB string theory \cite{WIT,TSE}.
The BPS states of the SYM theory preserving  1/2 of its supersymmetries
such as W-bosons, monopoles and dyons appear as $(p,q)$ strings
connecting two of the D3-branes in the IIB side.
The famous duality conjecture \cite{MON} of the ${\cal N}=4$ SYM 
theory which interchanges W-bosons and monopoles is obvious 
as the $SL(2,\Z)$ duality symmetry of IIB string theory.  

A pronged string \cite{PROSTR} can also end on the D3-branes.
It was conjectured that such configurations would appear as another
set of BPS states of the SYM theory preserving 1/4 of its
supersymmetries \cite{BER}.
The condition to preserve the 1/4 supersymmetries gives a set of field 
equations of the SYM theory, and its classical solutions were 
constructed \cite{CLSOUR,CLS}. 
The description of such BPS states in the full quantum treatment of 
the SYM theory is an open problem.

The M-theory gives another non-perturbative description of supersymmetric 
gauge theories.
The exact low energy effective lagrangian of the full quantum 
${\cal N}=2$ SQCD \cite{SEIWIT} can be described by an M5-brane embedded 
in a target space with one compactified direction \cite{WITTWO}.
The IIB string theory is related to the M-theory in the target space 
with two compactified directions, and a D3-brane is an M5-brane
wrapped in the two directions. Hence the ${\cal N}=4$ SYM theory may
be studied as the low energy dynamics of some parallel such M5-branes.
The 1/4 BPS states above should correspond to
M2-branes ending on the M5-branes and preserving the 1/4 of the 
supersymmetries of the M5-branes.

In this paper we shall investigate such M2-brane configurations. 
The cases without ends on M5-branes 
have already been discussed by several people \cite{MTH,KOL}.
A new thing in this paper is the presence of such ends.
Among other things, we shall show the existence of such M2-brane 
configurations. 
We shall also construct the configurations explicitly in some elementary
cases as examples. 
 
\section{BPS states in $\N=4$ SYM}
\label{BPS}
In this section we briefly review an M-theory description
of BPS states in four dimensional gauge theory, and
arrange them to formulate 1/4 BPS states in the $\N=4$ SYM theory,
 following \cite{HY,M,MTH}.

\subsection{Complex structures and BPS states}

Consider M-theory compactified on a torus with the modular parameter
$\tau$. $\N=4$ $U(N)$ SYM with the gauge coupling $\tau$
is obtained as an effective worldvolume theory
on $N$ parallel M5-branes wrapped on the torus.
We take the coordinate of space-time
 $(x^0,x^1,\cdots,x^9,x^{10})$ with the
identifications 
\begin{eqnarray}
  (x^{10},x^9)\sim (x^{10}+2\pi R,x^9)
  \sim (x^{10}+2\pi R\tau_1\,,x^9+2\pi R\tau_2),
\label{lat}
\end{eqnarray}
where $\tau_1$ and $\tau_2$ are real and imaginary part of $\tau$,
respectively. From now, we will abbreviate $R$ for simplicity,
although it is easy to recover the $R$ dependence from the dimensional
analysis.
 We also use complex variables
$v=x^4+x^5 i$ and $z=x^{10}+x^9 i$, which define a complex structure
$I$ of a 4-manifold $Q\equiv\R^2\times T^2=\{(v,z)\}$.
M5-branes are taken to be stretched along $\R^{1,3}\times T^2
=\{(x^0,\cdots,x^3,x^9,x^{10})\}$
directions, and the transverse positions
correspond to vacuum expectation values of scalar fields in the SYM
\footnote{ 
We take the vacuum expectation value of the self dual 2-form field on
each M5-brane to be zero.
}.
For our purpose, it is enough to consider the case, in which
all the M5-branes are placed with $(x^6,x^7,x^8)=(0,0,0)$.
Let $v_a$ ($a=1,2,\cdots,N$) be the position of $a^{\rm th}$ M5-brane
on the $v$-plane. We take $v_a$ to be distinct and
consider the Coulomb branch of the SYM, on which the gauge symmetry is 
broken to $U(1)^N$.
Note that the M5-branes are fixed by the equations $v=v_a$,
which define Riemann surfaces holomorphically embedded in $Q$
with respect to the complex structure $I$.

BPS-saturated states in M5-brane worldvolume theory are
obtained by M2-branes, whose boundaries lie on the M5-branes.
The M2-brane worldvolume is decomposed as $\R\times\Sigma$,
where $\R$ is the time axis and
$\Sigma$ is a Riemann surface embedded in $Q$.
 In $\N=2$ MQCD, it is known that
 $\Sigma$ must be holomorphically embedded
in $Q$ with respect to another complex structure $J$
which is orthogonal
 to $I$ (i.e. $IJ=-JI$) in order to saturate the BPS condition
\cite{FS,HY,M}.
Of course, we can apply this argument to the $\N=4$ case,
implying that if $\Sigma$ is holomorphically embedded
in $Q$ with respect to the complex structure $J$,
at least 4 supersymmetries are preserved.
What is different from the $\N=2$ case is that
the M5-branes, projected on $Q$, are also holomorphic with respect to
another complex structure $I'$ with holomorphic coordinates
$(\ol v,z)$. 
Since $Q$ is a flat 4-manifold,
there are complex structures $I,J,K,I',J',K'$ satisfying
\begin{eqnarray}
  &&IJ=-JI=K,\\
  &&I'J'=-J'I'=K',\\
  &&II'=I'I,~IJ'=J'I,~IK'=K'I,~JI'=I'J,~etc.
\end{eqnarray}
We can also apply the above argument replacing the complex structure
$I$ with $I'$, and find that if $\Sigma$ is holomorphically embedded
in $Q$ with respect to the complex structure $J'$,
another 4 supersymmetries are preserved.
In conclusion, in order to obtain 1/4 BPS states (1/2 BPS states),
 $\Sigma$ should be holomorphic with respect
 to either $J$ or $J'$ (both $J$ and $J'$).

The complex coordinates on $Q$, which are holomorphic
with respect to the complex structure $J$, are given as
\begin{eqnarray}
  z^1&=&\cos\alpha\,x^4+\sin\alpha\,x^5+x^{10}i,\\
  z^2&=&-\sin\alpha\,x^4+\cos\alpha\,x^5+x^9i,
\end{eqnarray}
where $\alpha$ parameterizes the freedom of choice of the complex
structure $J$.
In the following sections, we will set $\alpha=0$, rotating $v$-plane:
\begin{eqnarray}
  z^1&=&x^4+x^{10}i,\\
  z^2&=&x^5+x^9i.
\end{eqnarray}
It is useful to define single valued coordinates
\begin{eqnarray}
\label{s}
  s&=&\exp\left( z^1-\frac{\tau_1}{\tau_2}z^2\right)
,\\ 
\label{t}
  t&=&\exp\left(\frac{z^2}{\tau_2}\right)
\end{eqnarray}
which parameterize $Q$ globally.
In section \ref{sec:3}, we will search for
the Riemann surface $\Sigma$, which can be
 expressed as the zero locus of
a holomorphic function,
\begin{eqnarray}
  f(s,t)=0.
\end{eqnarray}

\subsection{Charges and BPS mass formula}

The boundaries of the M2-brane lie on the M5-branes
and couple with the M5-brane worldvolume theory
via the interaction
\begin{eqnarray}
\label{Sint}
  S_{\rm int}\sim\int_{\sR\times\partial\Sigma}B^+,
\end{eqnarray}
where $B^+$ is the self dual 2-form field on the M5-branes.
Now the M5-branes are wrapped on the torus and
$U(1)$ gauge fields in four dimension are related to $B^+$ as 
$A^a_{\mu}\sim B^{a+}_{10\,\mu}$ or $\wt A^a_{\mu}\sim
B^{a+}_{9\,\mu}$, where the superscript $a$ represents
 $a^{\rm th}$ $U(1)$ gauge field coming from the $a^{\rm th}$
M5-brane. As the field strength of $B^+$ is self dual, 
one can see that $A^a_{\mu}$ and $\wt A^a_{\mu}$ are ele-mag dual to
each other. If the homology class of $\partial\Sigma$ is
$n_e^a\alpha_a+n_m^a\beta_a$, where $\alpha_a$ and $\beta_a$ are
$\alpha$-cycle ($x^{10}$ direction) and $\beta$-cycle ($x^{9}$
direction) of the torus on the $a^{\rm th}$ M5-brane, (\ref{Sint})
implies
\begin{eqnarray}
  S_{\rm int}\sim(n^a_e+n^a_m\tau_1)\int_{\sR}A^a_{\mu}\,dx^{\mu}
  +n^a_m\tau_2\int_{\sR}\wt A^a_{\mu}\,dx^{\mu}.
\end{eqnarray}
{}From this, we can interpret $Q^a_e\equiv n^a_e+n^a_m\tau_1$
as the electric charges and $Q^a_m\equiv n^a_m\tau_2$ as the magnetic charges
of the BPS states.\cite{FS,MNS,S}

One of the significance of the above construction of the BPS states
is that we can easily compute mass of the BPS states using
the BPS mass formula. As shown in \cite{FS,HY}, mass of the BPS state
is given by a simple formula
\begin{eqnarray}
\label{mass}
  M=\left|\,\int_\Sigma\Omega\,\right|,
\end{eqnarray}
where $\Omega$ is a 2-form on $Q$, which is holomorphic with respect
to the original complex structure $I$ (or $I'$).
In our case, we have $\Omega=dv\wedge dz=d(v\,dz)$ and (\ref{mass}) can 
be expressed as
\begin{eqnarray}
\label{mass2}
  M=\left|\,\int_{\partial\Sigma}v\,dz\,\right|
  =\left|\,(n^a_e+n^a_m\tau)\,v_a\,\right|,
\end{eqnarray}
which is exactly what we expect from the field theory analysis
\cite{FH} or mass of the string web in type IIB string theory \cite{BER,BK}.
In fact, if we define electric and magnetic charge vectors as
\begin{eqnarray}
  \vec Q_E&=&Q^a_e\, (x^4_a,x^5_a),\\
  \vec Q_M&=&Q^a_m\, (x^4_a,x^5_a),
\end{eqnarray}
where $v_a=x^4_a+x^5_a i$, (\ref{mass2}) can be rewritten in the
more familiar form:
\footnote{If $\vec Q_E\times\vec Q_M$ turns out to be negative,
we should take the complex structure $I'$ to obtain (\ref{mass3})}
\begin{eqnarray}
\label{mass3}
  M=\sqrt{\,|\,\vec Q_E|^2+|\,\vec Q_M|^2+2\,|\,\vec Q_E\times\vec Q_M|\,}.
\end{eqnarray}

\section{Membrane configuration and electrodynamics}
\label{sec:3}

In this section, we shall discuss the configuration of the
holomorphically embedded 
M2-brane ending on M5-branes by an electric potential 
on the M2-brane worldsheet satisfying certain boundary conditions.

\subsection{General treatment}
\label{sec:general}
As has been discussed so far, the 1/4 supersymmetries are preserved if 
the embedding functions $v$ and $z$ of the M2-brane worldsheet $\Sigma$ 
are holomorphic functions on the Riemann surface $\Sigma$.
Locally this condition becomes the Cauchy-Riemann equations
\bea
d x^4 &=& * d x^{10}, \cr
d x^5 &=& * d x^{9},
\label{eq:cau}
\eea
where $*$ denotes the Hodge star operator on $\Sigma$.
The local integrability condition of \rf{eq:cau} 
is that the $x^4$ and $x^5$ be harmonic functions on $\Sigma$;
\be
\Delta x^{4}=\Delta x^{5}=0.
\label{eq:har}
\ee
Hence the functions $x^{4},x^{5},x^{9},x^{10}$ on the Riemann surface 
satisfying \rf{eq:cau} can be obtained by first 
solving \rf{eq:har} and then obtaining $x^{9},x^{10}$ by \rf{eq:cau}.
Boundary and global integrability conditions must also be 
considered. 

Let us consider an M2-brane ending on $n$  M5-branes.
In such a case, $\Sigma$ has $n$ connected components of its boundary 
$(\pt \Sigma)_a$ $(a=1,\cdots,n)$, where the M5-branes are attached.
Since an M5-brane has definite values of its locations $x^{4}$ and $x^{5}$,
the functions $x^4,x^5$ must take constant values at each boundary: 
\be
x^4(u)=x^4_a,\ x^5(u)=x^5_a,\ {\rm for} \ u\in(\pt \Sigma)_a,
\label{eq:boucon}
\ee
where $u$ denotes a complex coordinate on $\Sigma$, and $x^4_a$ and 
$x^5_a$ are the $a^{\rm th}$ M5-brane location.

The identification \rf{lat} of $x^9$ and $x^{10}$ gives the global 
constraints on the shifts of $x^9$ and $x^{10}$ under going along a 
one-dimensional cycle on $\Sigma$.
For boundaries, we have 
\bea
\int_{(\pt\Sigma)_a} dx^{10}&=&2\pi (n_e^a + \tau_1 n_m^a), \CR
\int_{(\pt\Sigma)_a} dx^9&=&2\pi \tau_2 n_m^a,  
\label{eq:quabou}
\eea
and, for the one-cycles $\alpha_i$ $(i=1,\cdots,2g)$ associated to the 
handles,
\bea
\int_{\alpha_i}dx^{10}&=&2\pi (N_e^i + \tau_1 N_m^i), \CR
\int_{\alpha_i}dx^9&=&2\pi \tau_2 N_m^i,
\label{eq:quahan}
\eea   
where $n_e^a,n_m^a,N_e^i,N_m^i$ are integers.
In the IIB string picture, the $n_e^a$ and $n_m^a$ are the two-form
charges of the strings ending on the D3-branes.
The $N_e^i$ and $N_m^i$ should be associated to the two-form
charges of the strings forming loops.
From (\ref{eq:quabou}), the conservation of the two-form
charges is derived : $\sum_a n^a_e=\sum_a n^a_m=0$. 

It is useful to introduce the language of electrodynamics to 
solve the above problem. Let us first discuss $x^4$ and $x^{10}$.
 
Firstly we can regard $x^4$ as an electric potential because of its
harmonicity. Then the boundary condition
\rf{eq:boucon} implies that the boundaries are conductors, on which
the electric potential must take constant values. 
The vector field $dx^4$ gives the electric vector field  associated to the
electric potential. From the Gauss law, the total electric charge in
a region can be evaluated by the line integral $\int *dx^4$ along its
boundary. Since $x^4$ is harmonic on $\Sigma$, the charges are located only
on the boundaries $(\pt\Sigma)_a$. Using \rf{eq:cau} and
\rf{eq:quabou}, the total electric charge on $(\pt\Sigma)_a$ is given by
\be
Q^a=\int_{(\pt\Sigma)_a} * dx^4 = 2\pi (n_e^a + \tau_1 n_m^a).
\label{eq:charge}
\ee
Thus $x^4$ is given by the electric potential generated by the electric 
charges $Q^a$ distributed on the conductors at $(\pt\Sigma)_a$.

With the above reinterpretation of the conditions \rf{eq:boucon} and 
\rf{eq:quabou}, the existence of $x^4$ for any configuration of 
$\Sigma$ with given charges $Q^a=2\pi(n_e^a + \tau_1 n_m^a)$
is intuitively obvious 
in case with vanishing genus. In this case, the values of $x^4$ at the
boundaries, i.e. the M5-brane locations, are given by a linear
function of the charges $Q^a$. 
The coefficients of the linear function is
determined by the configuration of $\Sigma$.
In fact they are invariant under the conformal transformation of 
$\Sigma$ because of the conformal invariance of the problem.
In a case with non-vanishing genus, the condition \rf{eq:quahan}
constrains further so that the electric fluxes associated to 
the handles must be quantized too. 
This constrains the possible configuration of $\Sigma$ in a discrete
manner. In fact, this is essential in
deriving the condition on the possible two-from charges of the 
pronged string configuration with one-loop. 
This issue will be discussed in the next subsection.

More rigorously, we can use some mathematical 
results on the Dirichlet first boundary value problem. 
There is a theorem \cite{F} implying that, for any given $x_a^4$, 
there exists a unique harmonic function $x^4$ on $\Sigma$ which
satisfies the boundary conditions
(\ref{eq:boucon})\footnote{Rigorously, some {\it local} 
conditions on the shapes of the boundaries must be satisfied, but
they seem irrelevant for our physical problem.}.
Since the harmonic function $x^4$ depends linearly on the boundary
values $x^4_a$,
the charges $Q^a$ depend linearly on $x_a^4$:
\be
Q^a=C^{ab} x^4_b + c^a,
\label{eq:linrel}
\ee
where the ``capacity'' $C^{ab}$ and $c^a$ are determined by the 
configuration of $\Sigma$.

To see the properties of $C^{ab}$ and $c^a$, first consider the
boundary condition that the $x_a^4$ takes an $a$-independent value, say
${x_0}^4$. Then the unique solution of $x^4$ is obviously the constant 
${x_0}^4$, and all the charges $Q^a=\int_{(\pt \Sigma)_a} * dx^4$  vanish. 
Thus we obtain $c^a=0$ and that $C^{ab}$ has an eigenvector
$x^4_a={x_0}^4$ with a vanishing eigenvalue.
On the other hand, if all the $Q^a$ vanish, $x_a^4$ takes an
$a$-independent value, say ${x_0}^4$. 
To prove this, suppose that $x_b^4$ is the maximum among all the
$x^4_a$'s. Then the maximum principle of the 
harmonic function implies $x^4 \leq  x_b^4$. Thus 
$Q^b= \int_{(\pt \Sigma)_b} * dx^4  \geq 0$. The equality holds only if 
$dx^4=0$ at $(\pt \Sigma)_b$. This boundary condition determines
uniquely $x^4= {x_b}^4={x_0}^4$ on $\Sigma$, and hence $x^4_a$ are
independent of $a$. Thus the linear relation \rf{eq:linrel} is invertible 
up to an arbitrary $a$-independent piece  ${x_0}^4$: 
\be
x^4_a=D_{ab} Q^b +  {x_0}^4=2\pi D_{ab} (n_e^b + \tau_1 n_m^b)  +  {x_0}^4.
\label{eq:rel4}
\ee
where $D_{ab}$ is determined by the configuration of $\Sigma$.
The $x^5_a$ can be just derived by substituting $Q^a=2\pi \tau_2
n_m^a$ in (\ref{eq:rel4}) with the same coefficients:
\be
x^5_a=2\pi \tau_2 D_{ab}  n_m^b +  {x_0}^5.
\label{eq:rel5}
\ee

In a case with non-vanishing genus, the electric fluxes associated to
handles \rf{eq:quahan} are also related linearly to the charges 
$n_e^a,n_m^a$ with coefficients determined by the configuration of $\Sigma$. 
Hence the quantization condition \rf{eq:quahan} constrains the possibility
of the coefficients, and so the configuration of $\Sigma$ is constrained
in a discrete manner.

The ``capacity'' matrix $C^{ab}$ is a symmetric matrix, as can be
found in a text book of electromagnetism. Thus we 
find\footnote{\rf{eq:iibcon} can be directly shown by noting that the
both sides are $\int_\Sigma * dx^4 dx^5$.} 
\be
x^5_a(n_e^a + \tau_1 n_m^a)=x^4_a\tau_2 n_m^a.
\label{eq:iibcon}
\ee
This equation agrees with  a necessary condition 
for that there exists a multi-pronged string connecting 
the D3-branes in the IIB string picture \cite{CLSOUR}.

\subsection{Examples}

In this subsection we shall explicitly construct the M2-brane
configurations in the following elementary cases, 
using the results in the previous subsection \ref{sec:general}. 
The first one corresponds to a tree-like three-pronged string two of 
the external strings end on D3-branes. The next one corresponds to a
one-loop multi-pronged string stretching to infinity. 

\subsubsection{Three-pronged string ending on D3-branes}

Here we shall explicitly construct the M2-brane configuration of a 
three-pronged string such that two of its ends are on the M5-branes 
while the other stretches to the infinity.
For simplicity, we take the type IIB coupling $\tau=i$ ($\tau_1=0,\tau_2=1$). 

\begin{figure}[htbp]
\begin{center}
\leavevmode
\epsfxsize=60mm
\epsfbox{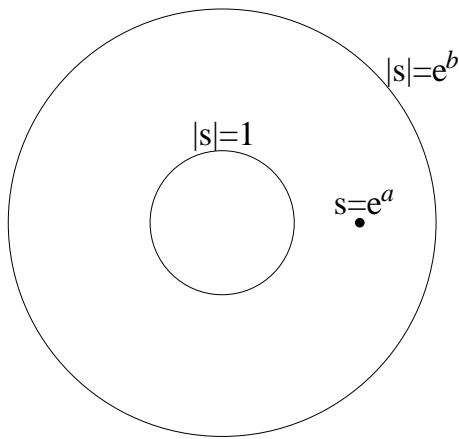}
\caption{The M2-brane worldsheet is mapped to the annulus region in 
the $s$-plane. The $(-1,0)$ and $(1,1)$ string ends at $x^4=0$ and $x^4=b$,
respectively, while the $(0,-1)$ string goes to $(x^4,x^5)=(a,-\infty)$.
}
\label{fig:annulus}
\end{center}
\end{figure}

The two-form charges of the strings we consider are
given by $(-1,0)$, $(0,-1)$ and $(1,1)$. 
Then the M2-brane worldsheet $\Sigma$ is mapped to an annulus region in
the $s$-plane as in fig.\ \ref{fig:annulus}.
Here we assumed that 
the $(-1,0)$ and $(1,1)$ strings end at $x^4=0$ and $x^4=b$,
respectively, while the $(0,-1)$ string goes to $(x^4,x^5)=(a,-\infty)$.
We may choose $s$ as the complex coordinate on $\Sigma$. 
Following the way in the preceding subsection \ref{sec:general},
the $x^5$ is given by 
an electric potential generated by a point-like charge of -1
at $s=e^a$ and a total charge of 1 on the conductor at 
$|s|=e^b$, while there is another conductor at $|s|=1$ with a total
charge zero.

\begin{figure}[htbp]
\begin{center}
\leavevmode
\epsfxsize=60mm
\epsfbox{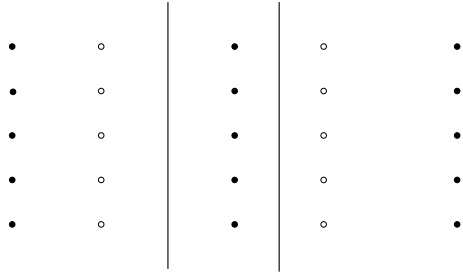}
\caption{The charges and the conductors in $\log(s)$ plane. The
original point-like charges are the small points between the
conductors represented by the two lines.  The others are the mirror
images. There are $\pm 1$ point-like charges represented in distinct ways.}
\label{fig:change}
\end{center}
\end{figure}

To obtain the electric potential, 
we may change the coordinate $s$ to $\log(s)$ as in 
fig.\ \ref{fig:change},
and apply the standard method of images.
By summing up all the contributions from the original point-like
charge and its images, 
we obtain 
\bea
t&=& s^{-a/b+1} \prod_{m,n=-\infty}^\infty 
\frac{\log(s)-a + 2n b + 2m \pi i}{\log(s)+a +2nb + 2m\pi i} \CR
&=& s^{-a/b+1}  \frac{\theta_1\left((\log(s)-a)/2b)|i\pi/b\right)}
{\theta_1\left((\log(s)+a)/2b)|i\pi/b\right)},
\eea
where $\theta_1$ is the Jacobi theta function and we put a 
point-like charge of $-a/b+1$ at $s=0$ to cancel the otherwise non-vanishing
total charge on the conductor at $|s|=1$. 

\subsubsection{Multi-pronged string with one loop}

To see how the quantization condition \rf{eq:quahan} appears,
we shall construct the M2-brane configuration associated to
a one-loop multi-pronged string the external strings of which 
stretch to infinity.      
In this case, the Riemann surface $\Sigma$ is a torus
with some punctured points.
The punctured points correspond to the ends of the infinitely
stretching external strings.
We parameterize $\Sigma$ with complex coordinate $u$ with
identifications $u\sim u+1$ and $u\sim u+\tau_\Sigma$,
where $\tau_\Sigma$ is the modular parameter of $\Sigma$.
This curve is embedded in $Q=\{(s,t)\}$ with
\begin{eqnarray}
  s=s\,(u),~~t=t\,(u),
\end{eqnarray}
where  $s\,(u)$ and $t\,(u)$ are some elliptic functions on the
worldsheet torus $\Sigma$.
{}From the definition of the parameters $s$ (\ref{s}) and $t$ (\ref{t}),
we know that the punctured points correspond
to the poles or zeros of the function $s\,(u)$ or $t\,(u)$.

Let us first discuss the function $s\,(u)$.
Suppose the two-form charges of the infinitely stretching external
strings are given by $(n_e^a,n_m^a)$ ($a=1,\cdots,n$).
Then the end of a string with a positive
$n_e^a$ should appear as an $n_e^a$-th order pole of $s$, while
one with a negative $n_e^a$ as an $-n_e^a$-th order zero.
Now suppose the ends are at $u=u_a$ on the torus $\Sigma$. 
Then, if and only if 
\be
\sum_{a=1}^n n_e^a u_a=-Q_1 - Q_2 \tau_{\Sigma}
\label{eq:elicon}
\ee
with some integers $Q_1$ and $Q_2$, 
there exists an elliptic function with the above desired property:
\be
s\,(u)=\exp (2(Q_1\eta_1+Q_2\eta_3)u) \prod_{a=1}^n \sigma(u-u_a)^{-n_e^a},
\ee
where the function $\sigma(u)$ is defined by
\be
\sigma(u)=\exp(\eta_1u^2)\theta_1(u|\tau_\Sigma)/\theta_1'(0|\tau_\Sigma)
\ee
and $\eta_1=\sigma'(1/2)/\sigma(1/2)$ and 
$\eta_3=\sigma'(\tau_\Sigma/2)/\sigma(\tau_\Sigma/2)$.

The construction of the function $t\,(u)$ is similar. The following 
quantization condition must be satisfied by the charges $n_m^i$ and 
some integers $q_1,q_2$: 
\be
\sum_{a=1}^n n_m^a u_a=-q_1 - q_2 \tau_\Sigma.
\label{eq:elicon2}
\ee
This equation gives further constraints on $u_a$.
Then
\be
t\,(u)=\exp (2(q_1\eta_1+q_2\eta_3)u) \prod_{a=1}^n \sigma(u-u_a)^{-n_m^a}.
\ee
There is a theorem that any two elliptic functions on a torus have 
an algebraic relation. Thus the torus coordinate $u$ can be eliminated,
and the M2-brane configuration should be given by the
zero locus of an algebraic function depending on the moduli of $\Sigma$:
\be
f_\Sigma(s,t)=0.
\ee

\begin{figure}[htbp]
\begin{center}
\leavevmode
\epsfxsize=60mm
\epsfbox{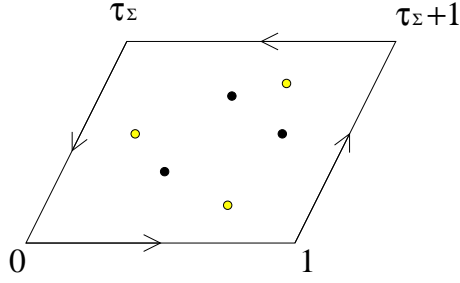}
\caption{The closed contour $C$ in the $u$-plane. The small points are 
the zeros and poles of $s$. }
\label{fig:torus}
\end{center}
\end{figure}

To see what \rf{eq:elicon}
 means, we consider the line integral
in fig.~\ref{fig:torus}:
\bea
\oint_C \log(s) du &=& i (x^{10}(u+1)-x^{10}(u))\tau_\Sigma -
i(x^{10}(u+\tau_\Sigma)-x^{10}(u))  \CR
&=& 2\pi i (N_1 \tau_\Sigma - N_2 ),
\eea
where the $N_1$ and $N_2$ are the electric fluxes crossing the two
one-cycles of the torus, respectively.
The other way to evaluate the integral is summing up the contributions 
from the zeros and poles of $s$:
\be
-\oint_C u d( \log(s)) = 2\pi i \sum_{a=1}^n  n_e^a u_a.
\ee
Thus \rf{eq:elicon} is just the quantization condition of the 
fluxes associated to the handles \rf{eq:quahan}, and gives constraints
on $u_a$.

\begin{figure}[htbp]
\begin{center}
\leavevmode
\epsfxsize=70mm
\epsfbox{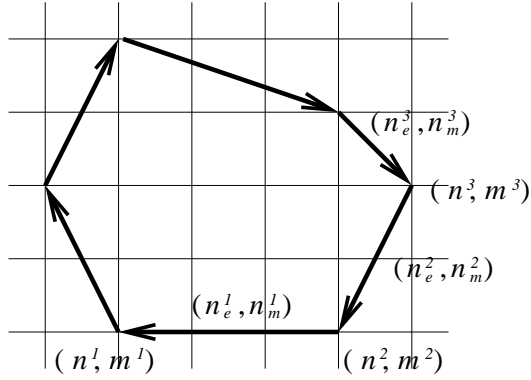}
\caption{The convex polygon
whose edges are given by the vectors $(n_e^a,n_m^a)$. }
\label{fig:polygon}
\end{center}
\end{figure}

We can also describe the conditions \rf{eq:elicon} and
\rf{eq:elicon2} graphically.
Suppose that $u_a$ are in the fundamental region:
\bea
  u_a=x_a+y_a\tau_\Sigma,~~~~(\,0\leq x_a,y_a \leq 1).
\eea
Then the conditions \rf{eq:elicon} and \rf{eq:elicon2} become
\bea
  \sum_{a=1}^n (n_e^a,n_m^a)\, x_a~\in\Z^2,
\label{xinzz}\\
  \sum_{a=1}^n (n_e^a,n_m^a)\, y_a~\in\Z^2.
\label{yinzz}
\eea
Since these two conditions are equivalent, it is enough to consider
\rf{xinzz} only.
We put the order of $x_a$ to satisfy 
$0\le x_1\le x_2\le\cdots\le x_n\le 1$,
and define
\bea
  t_1&=&x_1,\\
  t_i&=&x_i-x_{i-1}~~~(i>1),
\eea
which satisfy $t_i\ge 0$ and $\sum_{i=1}^n t_i\le 1$.
We also define
\bea
  (n^i,m^i)=\sum_{a=i}^n(n_e^a,n_m^a),
\eea
with $i=1,\cdots,n$.
Note that the charge conservation condition implies $(n^1,m^1)=(0,0)$.
Then we can rewrite the condition (\ref{xinzz}) as
\bea
  \sum_{i=1}^n (n^i,m^i)\, t_i~\in\Z^2,
\label{convex}
\eea
which implies that
there should exist a lattice point inside the convex hull of
the vertices $(n^i,m^i)$.
The condition \rf{convex} is most relaxed when
$(n^i,m^i)$ become vertices of a convex polygon
whose edges are given by the vectors $(n_e^a,n_m^a)$ as
 in fig.\ref{fig:polygon}.

Hence we conclude that the surface $\Sigma$ exists if and only if
there is an integer lattice point inside the convex polygon
whose edges are given by the vectors $(n_e^a,n_m^a)$. 
This condition is equivalent to that found in type IIB string
theory using the grid diagrams \cite{MNS,BK}. The modulus
represented by the size of the loop diagram in string junction
is now complexified and
corresponds to the modular parameter $\tau_\Sigma$ in the M-theory
description.

\section{Physical interpretation in type IIA string theory}
In the previous section, we used some technical methods
in electrodynamics to find out holomorphic surface $\Sigma$
with given winding numbers $(n^a_e,n^a_m)$.
The methods in electrodynamics was introduced with
a purely mathematical motivation,
and we did not care about the physical meaning.

In this section, we will try to make a physical interpretation of
the methods and answer the question why the electrodynamics
appeared in our problem.
It turns out that there is a natural interpretation from
the worldvolume gauge theory of D2-branes in the type IIA string theory.

Consider $n$ parallel  D2-branes stretched along $x^0,x^5,x^9$
directions. We will use the static gauge
\begin{eqnarray}
\label{static}
\sigma^0=x^0,~\sigma^1=x^5,~\sigma^2=x^9, 
\end{eqnarray}
where $(\sigma^0,\sigma^1,\sigma^2)$ are the worldvolume coordinates
of the D2-branes. The effective worldvolume theory is 3-dim $U(n)$ 
SYM which is obtained by dimensional reduction of 10-dim $\N=1$
$U(n)$ SYM. The bosonic part of the action is
\begin{eqnarray}
\label{D2}
  S=T\int d^3\sigma\Tr\left(
    -\frac{1}{4}F_{\mu\nu}F^{\mu\nu}+\frac{1}{2}D_\mu X^I D^\mu X^I
    +\frac{1}{4}[X^I,X^J]^2
    \right),
\end{eqnarray}
where $T$ is the D2-brane tension, and 
$X^I$ $(I=1,\cdots,4,6,\cdots,8)$ are adjoint scalar fields,
which represent the transverse fluctuations of the D2-branes.

Let us consider the BPS configuration of this system.
The energy is expressed as
\begin{eqnarray}
\label{energy}
  U=\frac{\,T\,}{2}\int d^2\sigma\Tr\left(
    \vec E^2+B^2+(D_0X^I)^2+(\vec DX^I)^2-\frac{1}{2}[X^I,X^J]^2
  \right),
\end{eqnarray}
where we have assumed that the fermion fields are zero.
Here we have defined electric and magnetic fields in 3-dim as
\begin{eqnarray}
  \vec E&=&(E_1,E_2)=(F_{01},F_{02}),\\
  B&=&F_{12}.
\end{eqnarray}
Introducing a unit vector $\eta^I$ in $\R^7$, we
have
\begin{eqnarray}
\label{energy2}
  U=\frac{\,T\,}{2}\int d^2\sigma\Tr\left(
    \,\left|\,\eta^I\vec E-\vec DX^I\,\right|^2
    +B^2+(D_0X^I)^2-\frac{1}{2}[X^I,X^J]^2
    +2\eta^I\vec E\cdot\vec DX^I
  \right).
\end{eqnarray}
Since the first four terms in the parenthesis in (\ref{energy2})
are positive definite, we obtain a bound for the energy
\begin{eqnarray}
\label{bound}
  U\geq T\,\eta^I\,Q_E^I,
\end{eqnarray}
where we have defined the charge vector
\begin{eqnarray}
  Q_E^I
  &=&\int d^3\sigma\Tr\left(\vec E\cdot\vec DX^I\right)\\
  &=&\int_{S^1_\infty} d\vec S\cdot\Tr\left(\vec E X^I\right).
\end{eqnarray}
Here we have used the Gauss's Law in the last equality.

The right hand side of (\ref{bound}) is maximized when
$\eta^I$ is proportional to the charge vector $Q^I_E$,
and then we obtain the Bogomol'nyi bound
\begin{eqnarray}
  U\geq T\,\|\,Q_E^I\|.
\end{eqnarray}
The BPS configurations, which saturate this bound, satisfy
the following equations:
\begin{eqnarray}
\label{BPSeq}
  \eta^I\vec E=\vec DX^I,~~B=0,~~D_0X^I=0,~~[X^I,X^J]=0.
\end{eqnarray}

Now consider the M-theory description of the D2-brane
configurations. We consider the case with
a single D2-brane ($n=1$ case).
As shown in \cite{T,BT,APPS}, the
M2-brane action can be obtained by performing a duality transformation 
of a worldvolume gauge field in the D2-brane action.
The scalar field corresponding to the fluctuations of the M2-brane
in $x^{10}$ direction is the dual of the worldvolume gauge field on
the D2-brane:
\begin{eqnarray}
\label{dual}
  F^{\mu\nu}=\epsilon^{\mu\nu\rho}\,\partial_{\rho}X^{10}.
\end{eqnarray}

Let us assume $X^I=0$ for $I=1,2,3,6,7,8$ as in the previous sections.
The BPS configurations (\ref{BPSeq}) are static configurations satisfying
\begin{eqnarray}
\label{BPS1}
  E_i=\partial_iX^4,~~B=0.
\end{eqnarray}
Hence, using (\ref{dual}), we obtain
\begin{eqnarray}
\label{CR}
  \epsilon_{0ij}\,\partial_{j}X^{10}&=&\partial_iX^4,\\
  \partial_0X^{10}&=&0.
\end{eqnarray}
(\ref{CR}), together with (\ref{static}), is nothing but the
Cauchy-Riemann equation
\begin{eqnarray}
  \frac{\partial X^4}{\partial x_5}
  =\frac{\partial X^{10}}{\partial x_9},
  ~~~\frac{\partial X^4}{\partial x_9}
  =-\frac{\partial X^{10}}{\partial x_5},
\end{eqnarray}
which imply that the M2-brane is holomorphically embedded in $Q$,
as explained in section \ref{BPS}.

In the previous section, we interpreted $X^4$ as the scalar
potential 
and the winding number $n_e^a$ on the boundary of
the membrane in $x^{10}$ direction 
as the electric charge in the 3-dim electrodynamics.
Now it is clear from (\ref{BPS1}) and (\ref{dual}) that
these interpretations can be naturally understood from the
electrodynamics of the D2-brane worldvolume gauge theory.

\vspace{.5cm}
\noindent
{\large\bf Acknowledgments}\\[.2cm]
We would like to thank K.\ Hashimoto, H.\ Hata and T.\ Kugo  
for valuable discussions.
We would like to thank also the organizers of the Summer Institute
`98, where we began the present work.
N.S. is Supported in part by Grant-in-Aid for Scientific
Research from Ministry of Education, Science and Culture
(\#09640346) and Priority Area: ``Supersymmetry and Unified Theory of 
Elementary Particles'' (\#707). 
S.S. is supported in part by Grant-in-Aid for JSPS fellows.

\newcommand{\JN}[4]{{\sl #1} {\bf #2} (#3) #4}
\newcommand{\andJ}[3]{{\bf #1} (#2) #3}
\newcommand{\AP}{Ann.\ Phys.\ (N.Y.)}
\newcommand{\MPL}{Mod.\ Phys.\ Lett.}
\newcommand{\NP}{Nucl.\ Phys.}
\newcommand{\PL}{Phys.\ Lett.}
\newcommand{\PR}{Phys.\ Rev.}
\newcommand{\PRL}{Phys.\ Rev.\ Lett.}
\newcommand{\PTP}{Prog.\ Theor.\ Phys.}
\newcommand{\NPP}{Nucl.\ Phys.\ Proc.\ Suppl.}
\newcommand{\JHEP}{J. High Energy Phys.}


\begin{thebibliography}{99}
\bibitem{WIT}
    E. Witten, \JN{\NP}{B460}{1996}{335}, hep-th/9510135.
\bibitem{TSE}
    A.A. Tseytlin, \JN{\NP}{B469}{1996}{51}, hep-th/9602064;
    M.B. Green and M. Gutperle, \JN{\PL}{B377}{1996}{28}, hep-th/9602077.
\bibitem{MON}
    C. Montonen and D. Olive, \JN{\PL}{B72}{1977}{117}.
\bibitem{PROSTR}
J.H. Schwarz, \JN{\NPP}{55B}{1997}{1}, hep-th/9607201;
O. Aharony, J. Sonnenschein and S. Yankielowicz,
\JN{\NP}{B474}{1996}{309}, hep-th/9603009.
\bibitem{BER}
    O. Bergman, \JN{\NP}{B525}{1998}{104}, hep-th/9712211.
\bibitem{CLSOUR}
K. Hashimoto, H. Hata and N. Sasakura, \JN{\PL}{B431}{1998}{303},
hep-th/9803127; hep-th/9804164.
\bibitem{CLS}
T. Kawano and K. Okuyama, \JN{\PL}{B432}{1998}{338}, hep-th/9804139;
K. Lee and P. Yi, \JN{\PR}{D58}{1998}{066005}, hep-th/9804174.
\bibitem{SEIWIT}
N. Seiberg and E. Witten, \JN{\NP}{B426}{1994}{19}, hep-th/9407087;
\JN{\NP}{B431}{1994}{484}, hep-th/9408099.
\bibitem{WITTWO}
E. Witten, \JN{\NP}{B500}{1997}{3}, hep-th/9703166.
\bibitem{MTH}
M. Krogh and S. Lee, \JN{\NP}{B516}{1998}{241}, hep-th/9712050;
Y. Matsuo and K. Okuyama, \JN{\PL}{B426}{1998}{294}, hep-th/9712070;
I. Kishimoto and N. Sasakura, \JN{\PL}{B432}{1998}{305}, hep-th/9712180.
\bibitem{KOL}
    B. Kol, hep-th/9705031;
    O. Aharony, A. Hanany and B. Kol, \JN{\JHEP}{01}{1998}{002},
    hep-th/9710116.
\bibitem{FS} A. Fayyazuddin and M. Spalinski,
  \JN{\NP}{B508}{1997}{219}, hep-th/9706087.
\bibitem{HY} M. Henningson and P. Yi, \JN{\PR}{D57}{1998}{1291},
  hep-th/9707251.
\bibitem{M} A. Mikhailov, \JN{\NP}{B533}{1998}{243}, hep-th/9708068.
\bibitem{MNS} A. Mikhailov, N. Nekrasov and S. Sethi,
  \JN{\NP}{B531}{1998}{345}, hep-th/9803142.
\bibitem{S} S. Sugimoto, \JN{\PTP}{100}{1998}{123},
  hep-th/9804114.
\bibitem{FH} C. Fraser and T. Hollowood, \JN{\PL}{B402}{1997}{106},
  hep-th/9704011.
\bibitem{BK} O. Bergman and B. Kol, hep-th/9804160.
\bibitem{T} P.K. Townsend, \JN{\PL}{B373}{1996}{68},
  hep-th/9512062.
\bibitem{BT} E. Bergshoeff and P.K. Townsend,
  \JN{\NP}{B490}{1998}{145}, hep-th/9611173.
\bibitem{APPS} M. Aganagic, J. Park, C. Popescu, and J.H. Schwarz,
  \JN{\NP}{B496}{1997}{215}, hep-th/9702133.
\bibitem{F} See for example, O. Forster, ``Lectures on Riemann
  Surfaces'', Springer-Verlag.
\end{thebibliography}
\end{document}